\documentclass[prb,twocolumn,floatfix,showpacs]{revtex4-1}
\usepackage{graphicx,color}
\usepackage{amsmath,amssymb,bm}

\definecolor{darkred}{rgb}{0.90,0,0}
\definecolor{darkgreen}{rgb}{0,0.60,.2}
\definecolor{darkblue}{rgb}{0,0,1}
\definecolor{grey}{cmyk}{0,0,0,0.25}
\definecolor{orange}{cmyk}{0,0.6,0.8,0}

\newcommand{\ket}[1]{\left|#1\right\rangle}

\newcommand{\ii}{\text{i}}
\newcommand{\dd}{\text{d}}

\begin{document}

\title{Controlling Dynamical Quantum Phase Transitions}

\author{D.\ M.\ Kennes,$^1$ D. Schuricht,$^2$ and C.\ Karrasch$^1$}

\affiliation{$^1$Dahlem Center for Complex Quantum Systems and Fachbereich Physik, Freie Universit\"at Berlin, 14195 Berlin, Germany}
\affiliation{$^2$Institute for Theoretical Physics, Center for Extreme Matter and Emergent Phenomena, Utrecht University, Princetonplein 5, 3584 CE Utrecht, The Netherlands}

\begin{abstract}
We study the dynamics arising from a double quantum quench where the parameters of a given Hamiltonian are abruptly changed from being in an equilibrium phase A to a different phase B and back (A$\to$B$\to$A). As prototype models, we consider the (integrable) transverse field Ising as well as the (non-integrable) ANNNI model. The return amplitude features non-analyticities after the first quench through the equilibrium quantum critical point (A$\to$B), which is routinely taken as a signature of passing through a so-called dynamical quantum phase transition. We demonstrate that non-analyticities after the second quench (B$\to$A) can be avoided and reestablished in a recurring manner upon increasing the time $T$ spent in phase B. The system retains an infinite memory of its past state, and one has the intriguing opportunity to control at will whether or not dynamical quantum phase transitions appear after the second quench. 
\end{abstract}

\maketitle


\section{Introduction}
In the last two decades we have witnessed an impressive surge in experimental advances pushing the frontier of realising and controlling (effectively) closed non-equilibrium quantum many-body systems.\cite{Bloch-08,ReichelVuletic11,Polkovnikov-11} This is of fundamental importance as the consequences of quantum many-body physics are observably real in these systems as well as of practical relevance as they might pave the way to quantum technologies in the future.

Two quantities that have recently attracted a tremendous amount of attention are the return amplitude\cite{Silva08}
\begin{equation}
G(t)=\left\langle\Psi_0\right|e^{-\ii Ht}\left|\Psi_0\right\rangle,
\label{eq:loschmidt}
\end{equation} 
and its related rate function
\begin{equation}
l(t)=-\frac{1}{L}{\rm ln}\left|G(t)\right|^2.
\label{eq:ratefunc}
\end{equation}
Here $\left|\Psi_0\right\rangle$ is some initial state, usually taken to be the ground state of some Hamiltonian $H_0$, while the subsequent time evolution is governed by a different Hamiltonian $H$; a setup that is referred to as a quantum quench.\cite{CalabreseCardy06} Intuitively, one might think of the square of the return amplitude as the probability of the ground state to return to itself under the time evolution with $H$.

As has been pointed out by Heyl \emph{et al.},\cite{Heyl-13} the rate function \eqref{eq:ratefunc} in the transverse-field Ising chain shows non-analytic behaviour at ``critical times" $t_n^*$ provided the quantum quench has crossed the quantum critical point, i.e., if the ground states of the Hamiltonians $H_0$ and $H$ belong to different zero-temperature phases. The appearance of these critical times signals the breakdown of a Taylor expansion in time. Heyl \emph{et al.} also pointed out the mathematical analogy of the non-analyticities in the rate function as well as the manifestation of an equilibrium phase transition in the usual free energy, and in doing so motivating the introduction of the term dynamical quantum phase transition (DQPT) for the former. 

One of the hallmarks of equilibrium quantum phase transitions is the inability to adiabatically connect the ground state of one phase to the ground state of the other phase (of different symmetry). Therefore, a non-analyticity in the ground state energy is routinely encountered when passing between the phases, irrespective of the path chosen to achieve this crossing. In contrast, the robustness of DQPTs is much less clear. The appearance of DQPTs often\cite{KS13,Heyl14,Canovi-14} but not always\cite{Hickey-14,AndraschkoSirker14,VajnaDora14} coincide with whether or not a quantum critical point separates $H$ and $H_0$. Nevertheless, the further study of DQPTs has attracted a lot of theoretical interest\cite{Kriel-14,Heyl15,VajnaDora15,SchmittKehrein15,Sharma-15,
JamesKonik15,HuangBalatsky16,Divakaran-16,BudichHeyl16,
AbelingKehrein16,Sharma-16,PS16,Bhattacharya17new,Zunkovic-16,Heyl17,KS17,
Weidinger-17,HeylBudich17,Piroli-17,Halimeh17new,Zauner17new,Homrighausen17new,Lang17new,Heyl-18,TrapinHeyl18,
Piroli-18} as well as successful efforts to realise DQPTs in ionic\cite{Jurcevic-17}  and atomic\cite{Flaeschner-18} systems in optical lattices. 

In this paper, we add to this debate a surprising flexibility in controlling DQPTs by performing double quenches (within the free transverse-field Ising chain and a non-integrable generalisation thereof). We elaborate on how the appearance of DQPTs can be tuned simply by increasing the time between the first and second quench. In particular, we show that the system can exhibit all four combinations of absence or presence of non-analyticities before and after the second quench, respectively, as is illustrated for double quenches in the transverse-field Ising chain in Fig.~\ref{fig:fancyplot}. This does not only suggest that the appearance of DQPTs is very fragile but also indicates an intriguing long-term memory of the system. With this fragility in mind and motivated by recent experiments,\cite{Jurcevic-17} we comment on the relation between non-analyticities in the rate function and the time evolution of the magnetisation. We find that the correspondence of zeros in the magnetisation to the critical times $t_n^*$, observed earlier for the transverse-field Ising model,\cite{Heyl-13} does not survive in the double quench setup (similarly to when integrability-breaking terms are included\cite{KS13} in a single quench setup). This provides further evidence that the correspondence found for a single quench in the free case seems to be accidental.   

The rest of this paper is organised as followed: Section~\ref{sec:setup} gives a general introduction to the physical systems studied, the observables calculated, and the methods used. Section~\ref{sec:results} summarises our main results about the controllability of DQPT in both the transverse-field Ising model and the axial next-nearest-neighbour Ising (ANNNI) chain. In Section~\ref{sec:magnetization} we analyse the connection between non-analyticities in the rate function and the magnetisation. Finally, in Section~\ref{sec:outlook} we close with a concluding summary.  

\section{Setup, model, and methods}\label{sec:setup}
\subsection{Setup}
We compute the return amplitude \eqref{eq:loschmidt}
and its corresponding rate function (\ref{eq:ratefunc}) for a time-dependent Hamiltonian $H(t)$ that models a double quantum quench
\begin{equation}
H(t) = \begin{cases} H_0, & t<0,\\ H_1, & 0\leq t\leq T, \\ H_2, & T < t. \end{cases}
\label{eq:Ht}
\end{equation}
As before, $\ket{\Psi_0}$ is the ground state of an initial Hamiltonian $H_0$.

\subsection{Model}
Specifically we consider the following one-dimensional Hamiltonian
\begin{equation}
H(\Delta,g)=-J\sum_i\bigl[\sigma_i^z\sigma_{i+1}^z+\Delta\sigma_i^z\sigma_{i+2}^z+g\sigma_i^x\bigr],
\label{eq:h}
\end{equation}
where $\sigma_i^{x,y,z}$ denote Pauli matrices acting at site $i$. We assume $J>0$ and $g\ge 0$, while $\Delta$ can be positive or negative. For $\Delta=0$, one recovers the transverse-field Ising chain, which can be mapped to a system of free fermions and hence be solved exactly. The Ising chain exhibits a quantum phase transition\cite{Sachdev99} at $g_c=1$, which separates a ferromagnetic (FM) phase for $g<1$ from a paramagnetic (PM) phase for $g>1$. In the thermodynamic limit, the FM possesses two degenerate ground states $\ket{\pm}$ with $\langle\sigma_i^z\rangle\neq 0$, while the PM ground state with $\langle\sigma_i^z\rangle=0$ is unique.

For finite next-nearest neighbour interactions $\Delta\neq0$, one obtains the ANNNI chain.\cite{Selke88,SuzukiInoueChakrabarti13} The model can be mapped to a system of interacting fermions with interaction strength $\propto\Delta$, which can no longer be solved exactly. The phase diagram of this model has been studied by several methods.\cite{Rujan81,PeschelEmery81,Allen-01,Beccaria-06,Beccaria-07,SelaPereira11} In addition to the FM and PM it also possesses two additional phases at large, repulsive values of the interaction $\Delta>1$.

For the rest of this paper, we will keep $J$ fixed; our double quench is thus entirely determined by the three pairs of the values $(\Delta_m,g_m)$, $m=0,1,2$, together with $H_m=H(\Delta_m,g_m)$.

\subsection{Analytical approach}\label{sec:analytics}
For the analytical approach we consider a chain of length $L$ with periodic boundary conditions on the spin variables, $\sigma_{L+1}^a=\sigma_1^a$. Furthermore we restrict ourselves to double quenches in the transverse-field Ising model ($\Delta_m=0$) for which exact results can be obtained. To this end, we  map the model to non-interacting fermions via a Jordan--Wigner transformation (see, e.g., Ref.~\onlinecite{Calabrese-12jsm1}, which we follow in our notation). In the fermionic language, the Hamiltonian can be diagonalised straightforwardly,
\begin{equation}
H(\Delta=0,g)=\sum_k\epsilon_k(g)\,\left(\eta_k^\dagger(g)\eta_k(g)-\frac{1}{2}\right),
\label{eq:Isingfermion}
\end{equation}
where
\begin{equation}
\epsilon_k(g)=2J\sqrt{1+g^2-2g\cos k},
\label{eq:Isingenergy}
\end{equation}
and $\eta_k^\dagger(g)$ and $\eta_k(g)$ are fermionic creation and annihilation operators. Depending on the filling fraction, the fermions fulfil either anti-periodic boundary conditions with the momenta quantised as half-integer multiples of $2\pi/L$, or periodic boundary conditions with the momenta quantised as integer multiples of $2\pi/L$. The anti-periodic case is usually referred to as Neveu-Schwarz (NS) sector while the periodic one is known as the Ramond (R) sector, respectively. 

The initial state for the double-quench protocol is given by the unique ground state of the fermionic model $\ket{0,g_0}$, which lies in the NS sector for any finite system. For $g_0>1$ this corresponds to the unique, PM ground state of the Ising model. We stress, however, that in the FM phase $0\le g<1$ the fermionic ground state corresponds to a superposition of the magnetic states $\ket{\pm}$.\cite{Calabrese-12jsm1}

The fermionic modes which diagonalise the Hamiltonian at different values of the transverse field are related via 
\begin{equation}
\begin{split}
\eta_k(g_1)=&\cos\frac{\theta_k(g_2)-\theta_k(g_1)}{2}\,\eta_k(g_2)\\
&+\ii\sin\frac{\theta_k(g_2)-\theta_k(g_1)}{2}\,\eta_{-k}^\dagger(g_2),\label{eq:Bogoliubov}
\end{split}
\end{equation}
where the Bogoliubov angle $\theta_k(g)$ is determined from
\begin{equation}
e^{\ii\theta_k(g)}=\frac{g-e^{\ii k}}{\sqrt{1+g^2-2g\cos k}}.
\end{equation}
Using the relations \eqref{eq:Bogoliubov} together with the fact that the initial state is the vacuum state, $\eta_k(g_0)\ket{0,g_0}=0$, the rate function for the return probability for times $t<T$ is found to be\cite{Silva08}
\begin{widetext}
\begin{equation}
l(t)=-\frac{1}{\pi}\int_0^\pi\dd k\,\ln\left|\cos^2\frac{\theta_k(g_1)-\theta_k(g_0)}{2}+\sin^2\frac{\theta_k(g_1)-\theta_k(g_0)}{2}\,e^{-2\ii\epsilon_k(g_1)t}\right|,
\label{eq:ratefunctionbefore}
\end{equation}
while for times $t>T$ we obtain
\begin{equation}
l(t)=-\frac{1}{\pi}\int_0^\pi\dd k\,\ln\left|A_k+B_k\,e^{-2\ii\epsilon_k(g_2)t}\right|+2\ln 2.
\label{eq:ratefunctionafter}
\end{equation}
The coefficients $A_k$ and $B_k$ depend on all quench parameters and are explicitly given by
\begin{eqnarray}
A_k&=&1+\cos\bigl[\theta_k(g_0)-\theta_k(g_1)\bigr]+\cos\bigl[\theta_k(g_0)-\theta_k(g_2)\bigr]+\cos\bigl[\theta_k(g_1)-\theta_k(g_2)\bigr]\nonumber\\
& &+\Bigl(1-\cos\bigl[\theta_k(g_0)-\theta_k(g_1)\bigr]+\cos\bigl[\theta_k(g_0)-\theta_k(g_2)\bigr]-\cos\bigl[\theta_k(g_1)-\theta_k(g_2)\bigr]\Bigr)e^{-2\ii\epsilon_k(g_1)T},\\
B_k&=&\Bigl(1+\cos\bigl[\theta_k(g_0)-\theta_k(g_1)\bigr]-\cos\bigl[\theta_k(g_0)-\theta_k(g_2)\bigr]-\cos\bigl[\theta_k(g_1)-\theta_k(g_2)\bigr]\Bigr)e^{2\ii\epsilon_k(g_2)T}\nonumber\\
& &+\Bigl(1-\cos\bigl[\theta_k(g_0)-\theta_k(g_1)\bigr]-\cos\bigl[\theta_k(g_0)-\theta_k(g_2)\bigr]+\cos\bigl[\theta_k(g_1)-\theta_k(g_2)\bigr]\Bigr)e^{-2\ii[\epsilon_k(g_1)-\epsilon_k(g_2)]T}.
\end{eqnarray}
\end{widetext}

\subsection{DMRG approach}\label{sec:dmrg}
In addition to the analytical approach discussed above, we employ the density matrix renormalisation group\cite{White92,white93,Schollwoeck11} (DMRG) to study the double-quench setup. The reason for this is two-fold: (1) The DMRG allows us to study quenches within the Ising chain that start from a polarised state, which is not a ground state of the fermionic model. Such quenches feature non-trivial dynamics of the magentisation and will be investigated in Sec.~\ref{sec:magnetization} in detail. (2) One can treat the ANNNI chain [$\Delta\neq0$ in Eq.~(\ref{eq:h})] which cannot be solved analytically; we will demonstrate that the picture described in Sec.~\ref{sec:generalresults} persists in such a non-integrable model.

At the technical level, we employ an infinite-system DMRG algorithm that is set up directly in the thermodynamic limit. We first determine the ground state using an evolution in imaginary time and then carry out a real time evolution to compute the rate function $l(t)$. The discarded weight is kept constant during the latter, which leads to a dynamic increase of the bond dimension. We performed every calculation using various different values of the discarded weight in order to ensure convergence. Further details of the numerical implementation can be found in Ref.~\onlinecite{KS13}.

\section{Results}\label{sec:results}
\subsection{General observations}\label{sec:generalresults}
Let us first recall\cite{Heyl-13} how DQPTs manifest for single quenches (i.e., $T=\infty$) within the Ising chain ($\Delta_m=0$). If the quench crosses the critical point $g=1$, the rate function (\ref{eq:ratefunc}) exhibits kinks in its time evolution, while such non-analytic behaviour is not observed if both $g_0$ and $g_1$ belong to the same phase (note, however, that for other models the appearance of DQPTs is no longer tied to the fact whether or not the quench crossed a critical point~\onlinecite{AndraschkoSirker14}).

For the double quench setup, we will demonstrate below that the appearance or absence of DQPTs does not only depend on the values of the quench parameters but also dramatically on the time $T$ between the first and the second quench. This entails a remarkable degree of controllability of the DQPTs. In fact, all four possible combinations for the absence or presence of kinks for times $t<T$ (after the first quench) and $t>T$ (after the second quench) can be realised. Strikingly, the existence of non-analytic behaviour in the rate function after the second quench can be tuned in a highly non-monotonic fashion, where in a recurring manner the DQPTs can be suppressed and re-instantiated by increasing $T$.

For future reference, we label the four cases mentioned above as follows: The rate function shows (i) no non-analyticities at all, (ii) no non-analyticities for $t<T$ but kinks for $t>T$, (iii) non-analyticities for $t<T$ but not for $t>T$, and (iv) kinks both for $t<T$ and $t>T$. The general observation that the appearance and absence of kinks can be tuned by varying $T$ is condensed in Fig.~\ref{fig:fancyplot}, which shows the critical times $t_n^*$ at which the rate function is non-analytic in dependence of the time $T$ for a typical set of parameters $g_0=1.5$, $g_1=0.5$, and $g_2=5.0$ in the Ising model (see Sec.~\ref{sec:analyticresults} for more details). Increasing $T$, we find recurring, discrete sets of lines of $t_n^*$ (solid lines in Fig.~\ref{fig:fancyplot}) which extend into the regime $t>T$. This illustrates that the appearance and vanishing of DQPTs after the second quench can be tuned by changing $T$. For a given value of $T$, the critical times $t_n^*$ at which rate function $l(t)$ shows kinks are determined by the crossing points of vertical lines in Fig.~\ref{fig:fancyplot} with the solid ones. Tokens of the classes (i)--(iv) defined above are thus, e.g., $TJ=0.5,1,2,3.5$, respectively. The recurring appearance and suppression of DQPTs after the second quench suggests an intriguing fragility of the concept (in contrast to the quite robust equilibrium quantum phase transitions) and gives rise to the high susceptibility to tuning outlined above. 

\subsection{Analytic results for the Ising chain}\label{sec:analyticresults}
\begin{figure}[t]
\includegraphics[width=0.9\linewidth,clip]{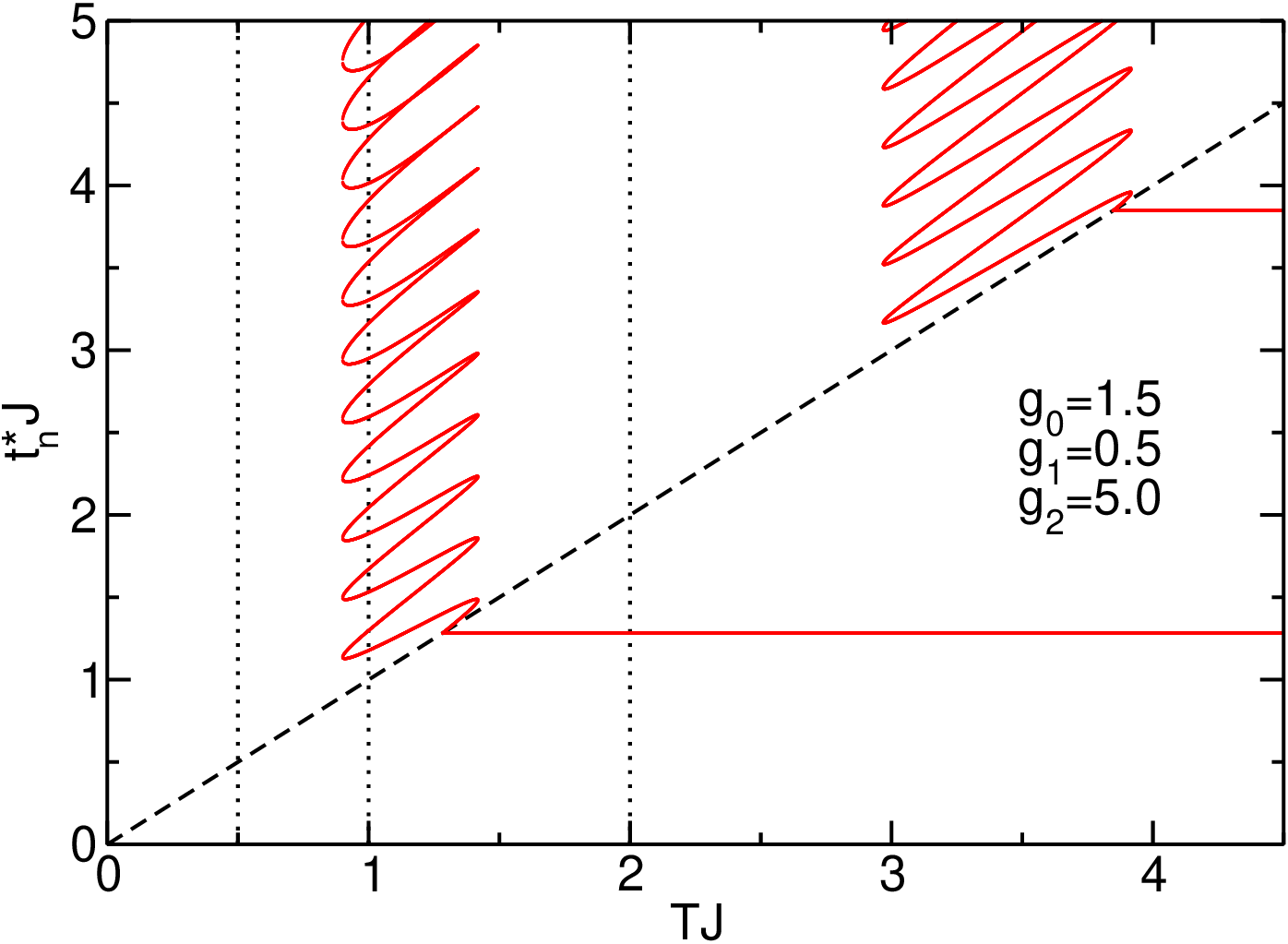}
\caption{(Colour online) Location of the critical times $t_n^*$ in the $t$-$T$-plane for a double quench between the PM$\to$FM$\to$PM phases of the transverse-field Ising chain (quench parameters $g_0=1.5$, $g_1=0.5$ and $g_2=5.0$). The dashed line marks the time $t=T$ at which the second quench is performed; the rate function $l(t)$ is shown explicitly in Fig.~\ref{fig:ising1}(a) for the quench times $TJ=0.5,1,2$ marked by dotted lines. We stress that kinks at times $t>T$ only occur for specific quench durations.}
\label{fig:fancyplot}
\end{figure}

The analytic results presented in Sec.~\ref{sec:analytics} allow us to obtain a complete understanding of the appearance of DQPTs for the transverse-field Ising chain. For times $t<T$, our setup is equivalent to the sudden quench protocol which was originally studied by Heyl~\emph{et al.}\cite{Heyl-13} who realised that the rate function $l(t)$ will show non-analyticities at specific times $t_n^*$ which are determined by a vanishing argument of the logarithm in Eq.~(\ref{eq:ratefunctionbefore}). This happens if the quench crosses the quantum critical point, and the times $t_n^*$ are located at\cite{Heyl-13} 
\begin{equation}
t_n^*=\frac{\pi}{2\epsilon_{k^*}(g_1)}(2n+1),\quad n\in\mathbb{N}_0.
\label{eq:tstarbefore}
\end{equation}
Here, the critical momentum $k^*$ is obtained from the condition
\begin{equation}
\cot^2\frac{\theta_{k^*}(g_1)-\theta_{k^*}(g_0)}{2}=1
\end{equation}
and explicitly given by $k^*=\arccos[(1+g_0g_1)/(g_0+g_1)]$. Depending on the value of $T$, one may thus observe a finite number of kinks before the second quench, as also shown in Fig.~\ref{fig:fancyplot}.

\begin{figure}[t]
\includegraphics[width=0.95\linewidth,clip]{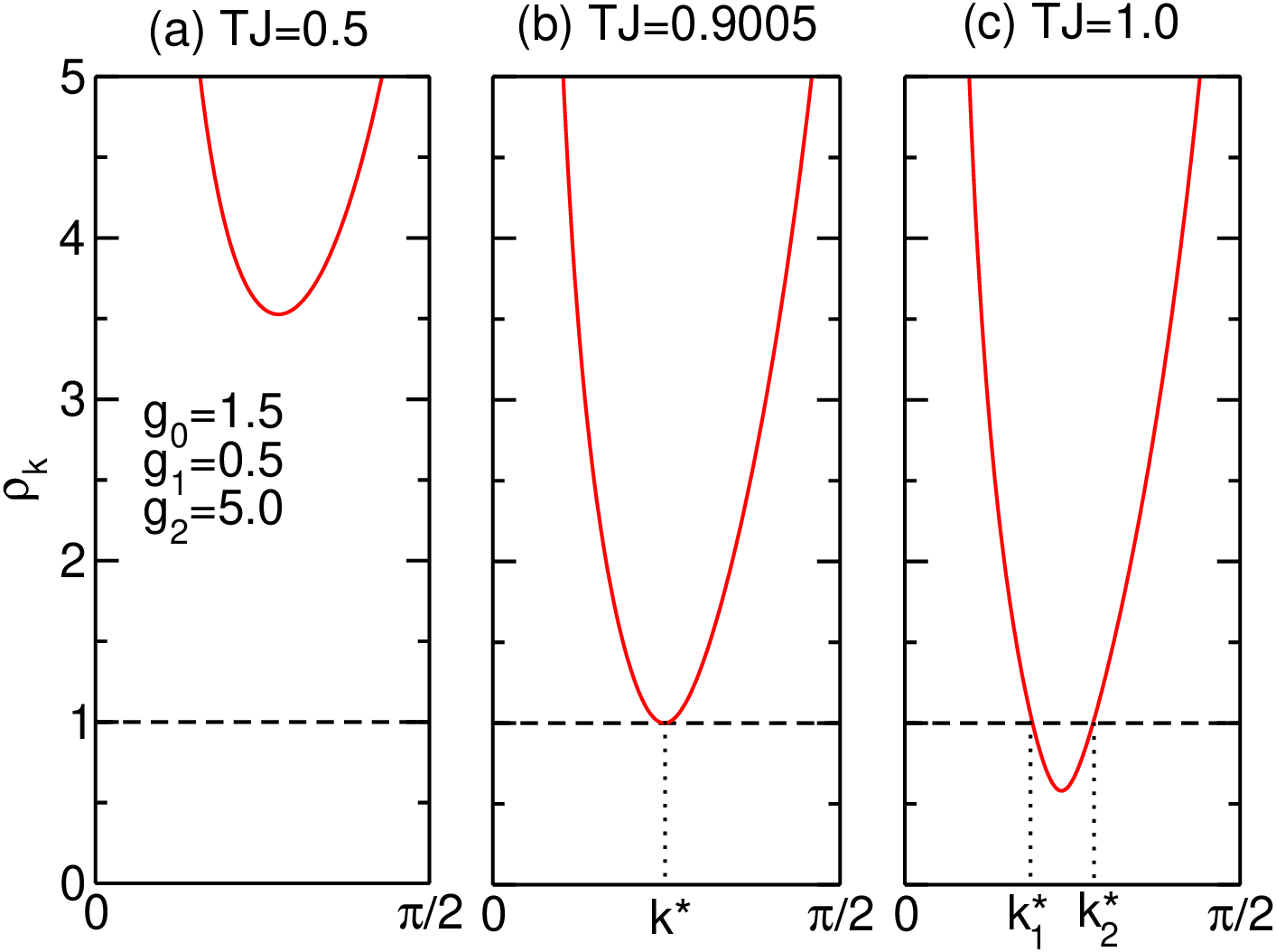}
\caption{(Colour online) Modulus $\rho_k=|A_k/B_k|$ for a double quench in the transverse-field Ising chain with quench parameters $g_0=1.5$, $g_1=0.5$ and $g_2=5.0$, and different quench times $T$. Generically we find either (a) no solution to \eqref{eq:condition2}, (b) one solution at $k=k^*$, or (c) two solutions $k=k_{1,2}^*$. Only the latter case results in critical times given by \eqref{eq:tstarlarget} at which the rate function shows non-analytic behaviour.}
\label{fig:rkplot}
\end{figure}
\begin{figure*}[ht]
\includegraphics[width=0.45\linewidth,clip]{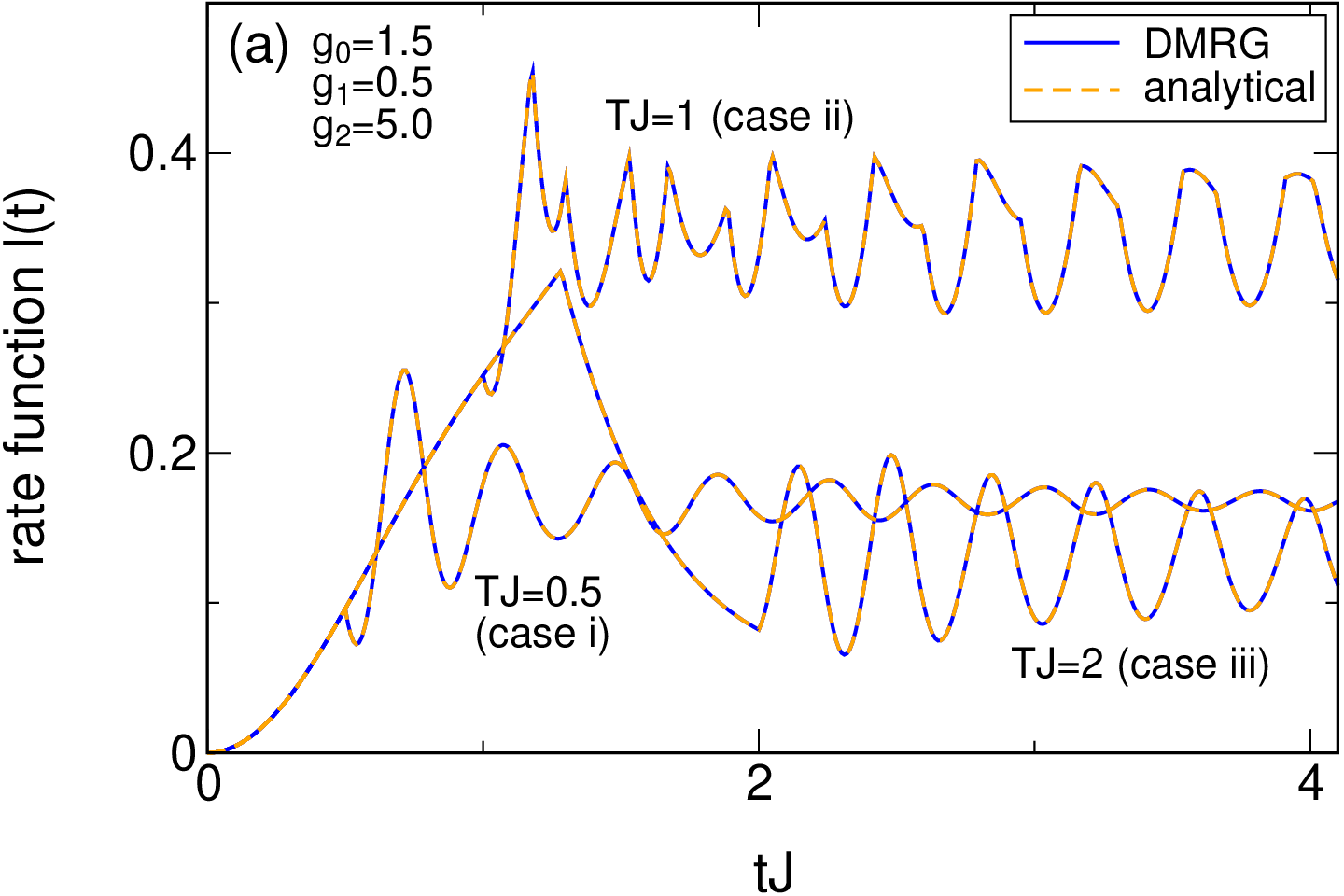}\hspace*{0.05\linewidth}
\includegraphics[width=0.45\linewidth,clip]{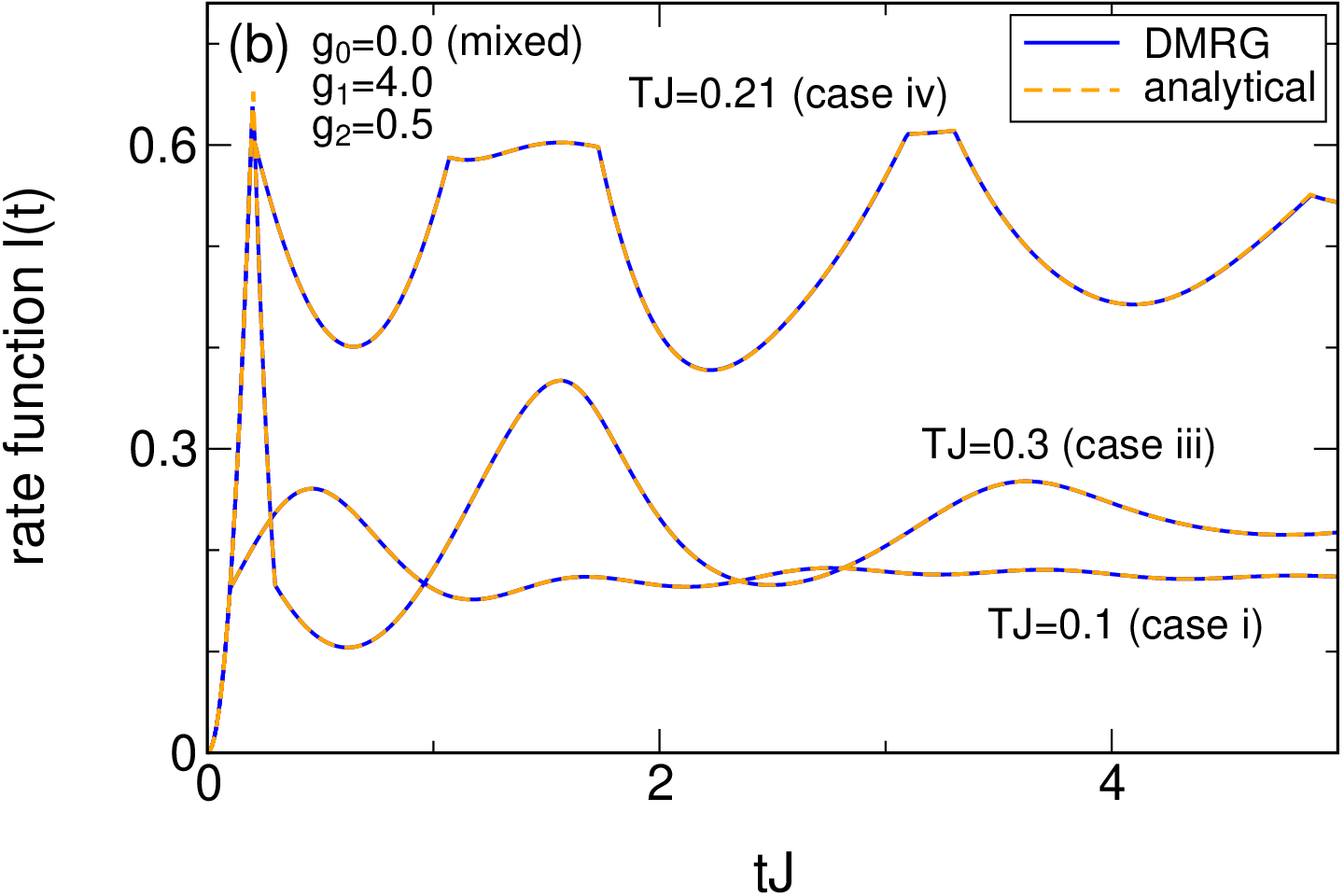}
\caption{(Colour online) Rate function $l(t)$ for a double quantum quench within the transverse-field Ising chain ($\Delta_m=0$) for different quench times $T$ and (a) quenches between the PM$\rightarrow$FM$\rightarrow$PM phases, (b) quenches between the FM$\rightarrow$PM$\rightarrow$FM phases starting from a mixed FM state. We compare the exact analytical results derived in Sec.~\ref{sec:analyticresults} with those obtained from a DMRG calculation. By varying $T$, one can systematically tune the appearance and suppression of DQPTs after the second quench. The different possible cases are discussed in Sec.~\ref{sec:generalresults}. }
\label{fig:ising1}
\end{figure*}
The situation becomes considerably more involved for times $t>T$ since Eq.~(\ref{eq:ratefunctionafter}) depends on all quench parameters $g_0$, $g_1$, and $g_2$ as well as the quench time $T$. To make our analysis more transparent, we will fix the values of the transverse field and discuss the dependence on $T$, also in light of the fact that this parameter can be directly controlled in experiments. As discussed above, non-analyticities in the rate function \eqref{eq:ratefunctionafter} will appear whenever the argument of the logarithm vanishes,
\begin{equation}
\frac{A_k}{B_k}+e^{-2\ii\epsilon_k(g_2)}=\rho_ke^{\ii\varphi_k}+e^{-2\ii\epsilon_k(g_2)}=0.
\label{eq:condition}
\end{equation}
The main difference to the case $t<T$ is that due to the time evolution until $t=T$, the coefficient $A_k/B_k=\rho_ke^{\ii\varphi_k}$ is now no longer real but in general complex. It is thus reasonable to introduce the modulus $\rho_k$ and the phase $\varphi_k$, for which the condition \eqref{eq:condition} implies
\begin{equation}
\rho_{k}=1.
\label{eq:condition2}
\end{equation}
For double quenches starting and ending in the same phase (e.g., $g_0,g_2>1$, $g_1<1$), we generically find one of the three cases shown in Fig.~\ref{fig:rkplot}: (a) There is no momentum for which Eq.~\eqref{eq:condition2} is satisfied; this is, for example, the case for quench times $TJ\lesssim 0.9$ or $2.97\lesssim TJ\lesssim 3.92$ for the parameters shown in Fig.~\ref{fig:fancyplot}. The time evolution of the rate function is then completely analytic for all $t>T$. (b) There is one critical momentum $k^*$ with $\rho_{k^*}=1$, while for all other momenta we have $\rho_k> 1$. As we will discuss below, there are also no non-analyticities in the time evolution in this case. (c) There are two critical momenta $k_1^*$ and $k_2^*$ at which Eq.~\eqref{eq:condition2} is satisfied. Close to these momenta, the function $\rho_k$ is linear, which implies non-analytic behaviour of the rate function at the times
\begin{equation}
t_{i,n}^*=\frac{\pi}{2\epsilon_{k^*_i}(g_2)}(2n+1)-\frac{\varphi_{k^*_i}}{2\epsilon_{k^*_i}(g_2)},
\label{eq:tstarlarget}
\end{equation}
where $i=1,2$ and $n\in\mathbb{N}_0$. The phase shifts $\varphi_{k_i^*}$ originate from the time evolution for $t<T$. We note that while the individual sets $\{t_{i,n}^*\}$, $i=1,2$, are periodic in time, due to the differing values of the prefactor $\pi/[2\epsilon_{k^*_i}(g_2)]$ the complete set of critical times $\{t_{1,n}^*\}\cup\{t_{2,n}^*\}$ is not periodic. In principle there may be more than two momenta at which \eqref{eq:condition2} is satisfied, each of them giving rise to a set of critical times determined by \eqref{eq:tstarlarget}.

The link between the cases (a)--(c) discussed here and the general cases (i)--(iv) introduced in Sec.~\ref{sec:generalresults} is as follows: Depending on whether or not the critical times \eqref{eq:tstarbefore} appear up to $T$, the cases (a) and (b) result in the general cases (i) or (iii). Similarly, case (c) leads to the cases (ii) or (iv).

Furthermore, the analytic result \eqref{eq:ratefunctionafter} allows us to analyse the behaviour of the rate function close to the critical times \eqref{eq:tstarlarget}. We expand the integrand around $k=k_i^*$ and $t=t_{i,n}^*$. As illustrated in Fig.~\ref{fig:rkplot}(c), $\rho_k$ is linear near $k_i^*$, $\rho_k\approx 1+a(k-k_i^*)$, $a\in\mathbb{R}$, thus we can approximate the rate function as follows:
\begin{eqnarray}
l(t)&\sim&-\frac{1}{2\pi}\int_0^\pi\dd k\,\ln\Bigl[a^2(k-k_i^*)^2+(2\epsilon_{k_i^*}(g_2)\delta t)^2\Bigr]\nonumber\\
&\sim&\delta t=|t-t_{i,n}^*|.
\label{eq:lineart}
\end{eqnarray}
This linear behaviour seems to be a general feature of DQPTs; it was previously observed after quenches across the quantum critical point in the transverse-field Ising model\cite{Heyl15} as well as quantum Potts chain.\cite{KS17}

Finally, let us comment on the case where there is precisely one critical momentum $k^*$ as shown in Fig.~\ref{fig:rkplot}(b). At this value, the modulus $\rho_k$ is no longer linear in $k-k^*$. Instead, we observe that when approaching the critical quench duration $T_\text{c}$ (given by $T_\text{c}J\approx 0.9005$ for the parameters of Fig.~\ref{fig:rkplot}) from above, the critical momenta $k_{1,2}^*$ and thus the times $t_{1,n}^*$ and $t_{2,n}^*$ approach each other, and eventually the kinks in the rate function simply disappear.

\subsection{Numerical results for the ANNNI chain}
We start by benchmarking our DMRG data for the Ising chain against the analytic results of Sec.~\ref{sec:analytics}. Fig.~\ref{fig:ising1} shows the rate function for two quenches starting from (a) the PM ground state, and (b) the mixed FM ground state that corresponds to the NS state in the fermionic language. By varying the quench time $T$, one can realise each of the different cases discussed in Sec.~\ref{sec:generalresults}. For example, for a quench starting in the mixed FM ground state [Fig.~\ref{fig:ising1}(b)], there are DQPTs for $TJ=0.1$ (case i), for $TJ=0.19$, kinks appear only for $t>T$ (case ii; not shown in the figure), for $TJ=0.3$, there are kinks only for $t<T$ (case iii), and for $TJ=0.21$, kinks appear for both $t<t$ and $t>T$ (case iv). In all cases, the DMRG data agree perfectly with the exact result.

\begin{figure}[t]
\includegraphics[width=0.9\linewidth,clip]{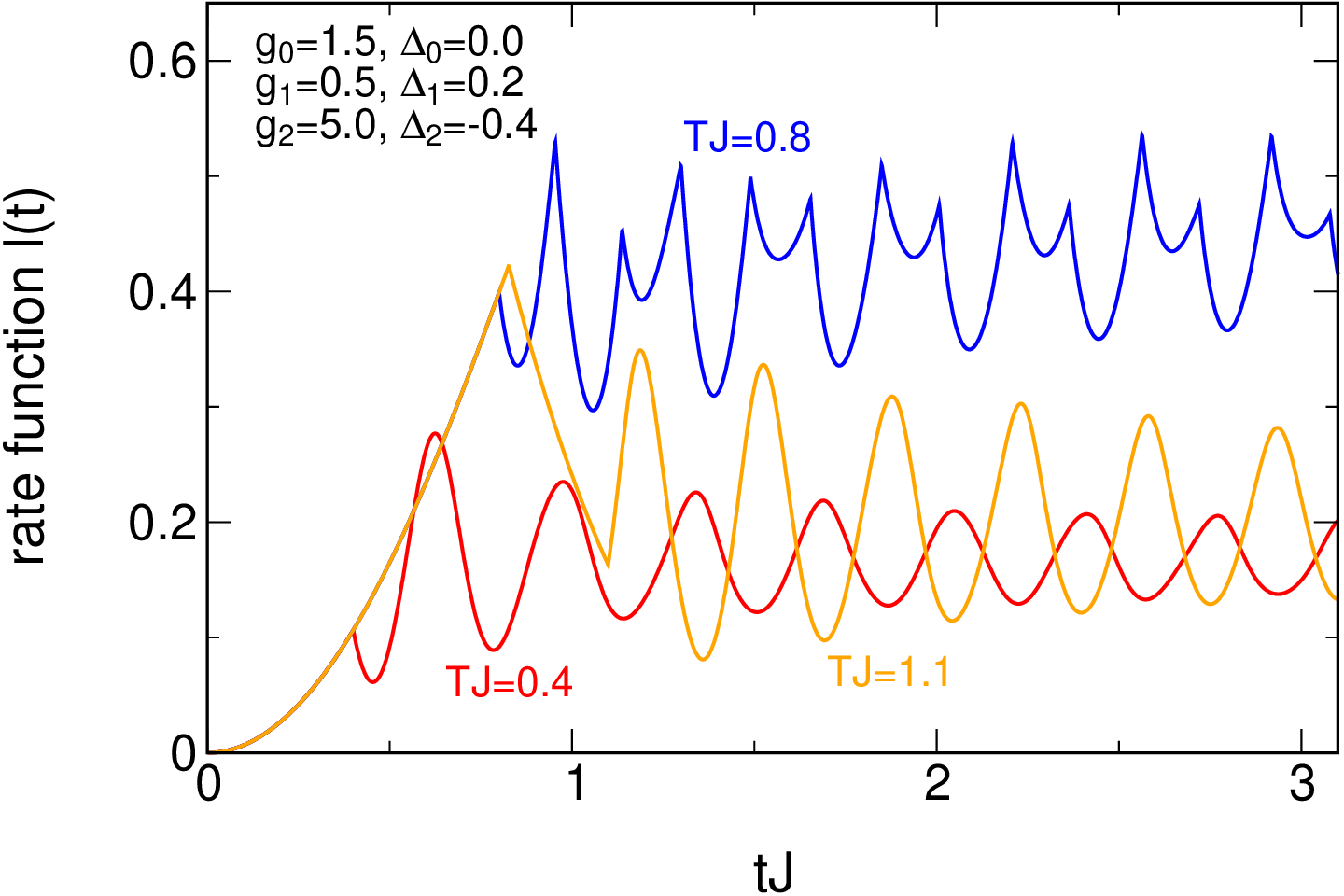}
\caption{(Colour online) Rate function for a double quench PM$\rightarrow$FM$\rightarrow$PM with different quench times $T$ for the ANNNI model. The results were obtained using DMRG. }
\label{fig:annni}
\end{figure}

Our general results, which we discussed in Sec.~\ref{sec:generalresults}, were mainly based on the analytic solution of the transverse-field Ising model. It is important to show that the main conclusions  are robust against breaking the integrability of this model and are therefore expected to hold in generic quantum many-body systems. To this end, in Fig.~\ref{fig:annni} we report results on the ANNNI model with finite $\Delta$, which, to the best of our knowledge, is not integrable. We show that in complete analogy to the free (integrable) case, the behaviour of the rate function $l(t)$ can be flexibly controlled by changing $T$; we explicitly demonstrate the appearance of the three cases: (i) no non-analyticities ($TJ=0.4$, red solid curve online), (ii) no non-analyticities for $t<T$ but kinks for $t>T$ ($TJ=0.8$, blue solid curve online),  and (iii) non-analyticities for $t<T$ but no kinks for $t>T$ ($TJ=1.1$, orange solid curve online). While we cannot rule out that a different phenomenology emerges at larger times inaccessible to the DMRG, the data of Fig.~\ref{fig:annni} indicate that flexible control of the appearance of DQPTs is possible even in non-integrable models; one can suppress and re-establish DQPTs at will.  

\section{Relation between DQPTs and magnetisation}\label{sec:magnetization}
By a remarkable experimental effort, the authors of Ref.~\onlinecite{Jurcevic-17} succeeded in directly measuring the rate function in a string of up to 10 Calcium ions, which are used to simulate long-ranged Ising models. Although a system of 10 ions is admittedly small, the work established a connection between the theory of DQPTs and the non-equilibrium physics expected in real quantum simulators. In Ref.~\onlinecite{Jurcevic-17} also the magnetisation was addressed for a quench starting from a FM polarised state, which is arguably a more natural quantity than the rate function. It was shown that the times where the magnetisation vanishes are tied to the critical times $t_n^*$ where kinks in the rate function show up. In Ref.~\onlinecite{Heyl-13}, a similar connection was observed for the transverse field Ising chain in the thermodynamic limit.   In contrast, it was previously demonstrated that such a direct relation does \emph{not} carry over to the non-integrable case such as the ANNNI model.\cite{KS13} This begs the question whether or not such a relationship between zeros in the magnetisation and the critical times in the rate function exists for more general setups. 

\begin{figure}[t]
\includegraphics[width=0.9\linewidth,clip]{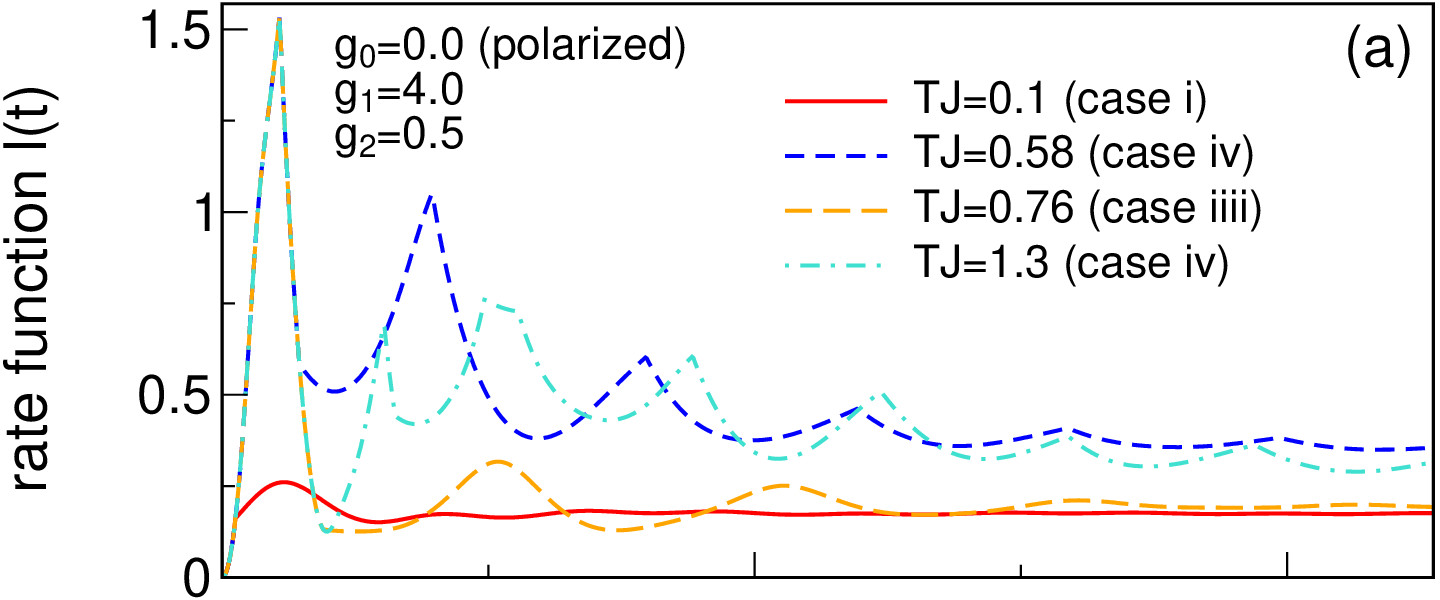}
\includegraphics[width=0.9\linewidth,clip]{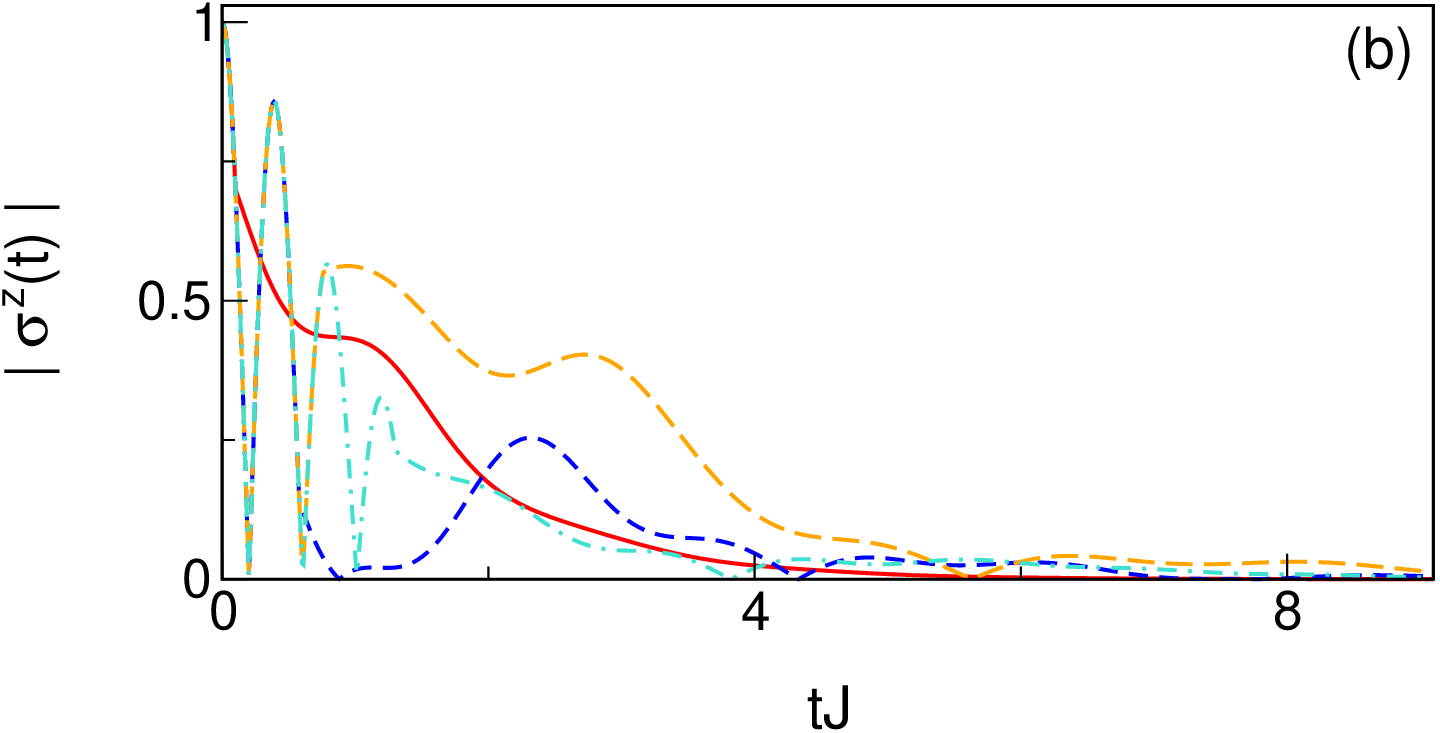}
\caption{(Colour online) (a) The same as in Fig.~\ref{fig:ising1}(b), but starting from a polarised FM state. (b) Behaviour of the magnetisation $|\langle\sigma^z(t)\rangle|$ during this type of quench. The data was computed using DMRG. We show that the times at which the magnetisation vanishes are in general not tied to the critical times $t_n^*$ where kinks in the rate function show up (in contrast to the transverse field Ising chain)}
\label{fig:ising2}
\end{figure}

For the double quench, we observe that this is \emph{not} the case (similarly to what is found for single quenches in non-integrable models), suggesting that the correspondence is not robust (even for free models). In Fig.~\ref{fig:ising2}, we explicitly compare the rate function and the magnetisation for a double quench starting from a FM polarised state of the transverse-field Ising model. The data was obtained using the DMRG method. One can explicitly see that the kinks in $l(t)$ at times $t>T$ after the second quench are in general  \emph{not} related to the zeros in the magnetisation. This is most prominent in the $TJ=0.58$ curve (dashed blue curve online), where the rate function for times $t>T$ shows repeated kinks, while the magnetisation vanishes at a completely different time scale.

\section{Conclusion}\label{sec:outlook}
In this paper, we have studied the phenomenon of DQPTs after double quantum quenches A$\to$B$\to$A between two equilibrium phases A and B. We have calculated the rate function analytically for the free transverse-field Ising chain. By varying the time $T$ spent in phase B, one can control at will and in a recurring manner whether or not DQPTs occur after the second quench. All four possible combinations of the appearance and absence of non-analyticities before and/or after the second quench can be realised if $T$ is tuned. A similar picture emerges using finite time DMRG numerics for the ANNNI model which is a non-integrable generalisation of the Ising chain. Moreover, we demonstrated that even for the free transverse-field Ising chain there is no relationship between the critical times after the second quench and the zeros of the magnetisation. In conclusion, our results show that the appearance of DQPTs is very fragile against the details of the quench setup.

\section*{Acknowledgments}
 C.K. and D.K. are supported by the DFG via the Emmy-Noether program under KA 3360/2-1. D.S. is member of the D-ITP consortium, a program of the Netherlands Organisation for Scientific Research (NWO) that is funded by the Dutch Ministry of Education, Culture and Science (OCW). D.S. was supported by the Foundation for Fundamental Research on Matter (FOM), which is part of the Netherlands Organisation for Scientific Research (NWO), under 14PR3168.

\end{document}